\documentclass[amsmath,amssymb,aps, physrev, showpacs, superscriptaddress,notitlepage]{revtex4-1}

\usepackage{graphicx}
\usepackage{dcolumn}
\usepackage{bm}
\usepackage{braket,amsmath,amssymb,bm}
\usepackage{color}
\usepackage{siunitx}
\usepackage{float}
\usepackage{hyperref}
\usepackage{lineno}
\usepackage{multirow}
\usepackage{tabularx}
\usepackage{enumitem}
\DeclareSIUnit\molar{\mole\per\cubic\deci\metre}
\DeclareSIUnit\Molar{M}

\begin{document}

\title{Wide-field fluorescent nanodiamond spin measurements toward real-time large-area intracellular quantum thermometry}

\author{Yushi Nishimura}
 \affiliation{Department of Chemistry, Osaka City University, Sumiyoshi-ku, Osaka, 558-8585, Japan}
 
 \author{Keisuke Oshimi}
 \affiliation{Department of Chemistry, Osaka City University, Sumiyoshi-ku, Osaka, 558-8585, Japan}

\author{Yumi Umehara}
 \affiliation{Department of Chemistry, Osaka City University, Sumiyoshi-ku, Osaka, 558-8585, Japan}
 
 \author{Yuka Kumon}
\affiliation{Department of Biomolecular Engineering, Graduate School of Engineering, Nagoya University, Nagoya 464-8603, Japan}
 
 \author{Kazu Miyaji}
\affiliation{Department of Biomolecular Engineering, Graduate School of Engineering, Nagoya University, Nagoya 464-8603, Japan}
 
 \author{Hiroshi Yukawa}
\affiliation{Department of Biomolecular Engineering, Graduate School of Engineering, Nagoya University, Nagoya 464-8603, Japan}
\affiliation{Institute of Nano-Life-Systems, Institutes of Innovation for Future Society, Nagoya University, Nagoya 464-8603, Japan}
\affiliation{Institute of Quantum Life Science, National Institutes for Quantum and Radiological Science and Technology, Chiba, Japan.}
 
 \author{Yutaka Shikano}
 \affiliation{Quantum Computing Center, Keio University, 3-14-1 Hiyoshi, Kohoku, Yokohama, 223-8522, Japan}
 \affiliation{Institute for Quantum Studies, Chapman University, 1 University Dr., Orange, CA 92866, USA}
 
 \author{Tsutomu Matsubara}
 \affiliation{Department of Anatomy and Regenerative Biology, Graduate School of Medicine, Osaka City University, Osaka 545-8585, Japan}
 
\author{Masazumi Fujiwara}
 \email{masazumi@osaka-cu.ac.jp}
 \affiliation{Department of Chemistry, Osaka City University, Sumiyoshi-ku, Osaka, 558-8585, Japan}
 
 \author{Yoshinobu Baba}
\affiliation{Department of Biomolecular Engineering, Graduate School of Engineering, Nagoya University, Nagoya 464-8603, Japan}
\affiliation{Institute of Nano-Life-Systems, Institutes of Innovation for Future Society, Nagoya University, Nagoya 464-8603, Japan}
\affiliation{Institute of Quantum Life Science, National Institutes for Quantum and Radiological Science and Technology, Chiba, Japan.}

\author{Yoshio Teki}
 \affiliation{Department of Chemistry, Osaka City University, Sumiyoshi-ku, Osaka, 558-8585, Japan}

\begin{abstract}
In this study, we analyze the operational process of nanodiamond (ND) quantum thermometry based on wide-field detection of optically detected magnetic resonance (ODMR) of nitrogen vacancy centers, and compare its performance with that of confocal ODMR detection.
We found that (1) the thermometry results are significantly affected by the shape and size of the camera region of interest (ROI) surrounding the target NDs and that (2) by properly managing the ROI and acquisition parameters of the camera, a temperature precision comparable to confocal detection in living cells can be obtained by wide-field ODMR. 
Our results are significant to the development of camera-based real-time large-area quantum thermometry of living cells.
\end{abstract}
\maketitle

\section{Introduction}

Sub-cellular thermometry has great potential in studying molecular mechanisms of temperature-related biological phenomena, such as the variation of cell-death types in photothermal cancer therapy~\cite{shah2008photoacoustic,zhu2016temperature,zhang2018temperature}, cellular thermotaxis~\cite{TCR12829,paulick2017mechanism,sekiguchi2018thermotaxis,Sagvolden471,Deman167}, and cellular level thermogenesis~\cite{kiyonaka2013genetically,okabe2012intracellular,yang2017measurement,doi:10.1002/anie.201915846,chretien2018mitochondria}.
Fluorescence-based nanodiamond (ND) quantum thermometry is a novel opto-microwave hybrid technique that can probe sub-cellular temperatures with distinct photo-stability~\cite{doi:10.1098/rsos.190589,reineck2017bright,mochalin2012properties,doi:10.1021/acsphotonics.5b00732}, various functionalized surface~\cite{lin2015protein,sotoma2018highly,doi:10.1002/anie.201905997} and ultra-low cytotoxicity~\cite{mohan2010vivo,simpson2017non,hsiao2016fluorescent,haziza2017fluorescent,doi:10.1021/jp066387v}.
The technique is based on optically detected magnetic resonance (ODMR) of nitrogen vacancy (NV) color defect centers in diamonds, where the fluorescence of NDs is measured under microwave irradiation to detect the fluorescence decrease at an electron spin resonance of about 2.87 GHz~\cite{doherty2013nitrogen,schirhagl2014nitrogen,petrini2020quantum,Levine2019principles}.
The ODMR frequency is temperature dependent, and one can probe the temperature of NDs by measuring the ODMR frequency shift~\cite{kucsko2013nanometre,PhysRevLett.104.070801,chen2011temperature,PhysRevX.8.011042,liu2019coherent,tzeng2015time,Toyli8417,doi:10.1021/acs.nanolett.8b00895,neumann2013high,clevenson2015broadband,PhysRevB.91.155404}. 
Biological applications of this method have been demonstrated for various cultured cells, including fibroblasts~\cite{kucsko2013nanometre}, neurons~\cite{simpson2017non}, and stem cells~\cite{yukawa2020quantum}.
It has been recently applied to \textit{in-vivo} nematode worms for studying physiological thermogenesis~\cite{fujiwara2019realtime} and the temperature dependence of embryogenesis~\cite{choi2019}. These demonstrations have proven the usefulness of ND quantum thermometry.

To further integrate this technique into biological thermometry, real-time operations and large-area probing are required. 
ND quantum thermometry is mainly classified into two types depending on the fluorescence detectors used, i.e., photon-counter-based confocal detection and camera-based wide-field detection.
Confocal detection is a point-by-point ND probing technique, which is relatively slow for working with multiple NDs, but suitable for implementing fast timing control of photon detection and microwave pulsing~\cite{kucsko2013nanometre,tzeng2015time,yukawa2020quantum,fujiwara2019realtime,choi2019,PhysRevX.8.011042}.
The measurement characteristics of confocal detection, such as noise characters, photon flux dependence, and measurement artifacts~\cite{fujiwara2020arxiv}, have been intensively studied over the years~\cite{kucsko2013nanometre,tzeng2015time,choi2019,fujiwara2019realtime}, 
Wide-field detection offers parallel probing of multiple NDs, which has recently been introduced in thermometry~\cite{simpson2017non,sekiguchi2018fluorescent,sotoma2018enrichment}; however, its measurement characteristics as thermometry have not yet been investigated comprehensively. 
Considering that the only difference between these two methods is the fluorescence detectors, knowledge of the confocal-ODMR-based quantum thermometry can be used to develop wide-field detection once the link between these two methods is established. 

In this study, we analyze wide-field ND quantum thermometry in relation with confocal detection.
Furthermore, we clarify the technical points that can specifically affect the temperature measurements in wide-field detection, including background fluorescence and selection of binning regions around target NDs. 
By properly managing these technical points, we demonstrate that wide-field ODMR detection can provide thermometric results in living cells comparable with confocal detection.
We discuss the possibility of realizing large-area quantum thermometry operable in real-time, which should be an important technological milestone of quantum sensor applications to biology.  

\section{Materials and Methods}
\subsection{Materials}
The NDs were purchased from Ad\'{a}mas Nanotechnologies (Raleigh, VA, USA). 
E-MEM with L-glutamine, phenol red, sodium pyruvate, non-essential amino acids, sodium bicarbonate (1,500 mg/L), Dulbecco’s phosphate-buffered saline (D-PBS(-)), and 4\% paraformaldehyde phosphate buffer solution were purchased from Fujifilm Wako Pure Chemical Corporation (Tokyo, Japan). 
D-MEM and fetal bovine serum (FBS) were purchased from Thermo Fisher Scientific (Waltham, MA, USA).
Collagenase Type I was purchased from Koken Co. (Tokyo, Japan).
A cell counting plate was purchased from Fukae Kasei Co. (Hyogo, Japan). 
Acti-stain 488 phalloidin was purchased from Cytoskeleton (Denver, CO, USA). 
Hoechst 33342 was purchased from Thermo Fisher Scientific (Tokyo, Japan). 
Fluoromount was purchased from Diagnostic BioSystems (Pleasanton, CA, USA).

\subsection{Cell preparation and ND labeling for super-resolution microscopy}
HeLa cells were incubated with the NDs (250 $\si{\ug}$/mL) on glass-based dishes in a transduction medium (D-MEM containing 10\% FBS and 100 U/mL penicillin/streptomycin) at 37$\si{\degreeCelsius}$. 
After a 24-h incubation, the cells were washed two times using the PBS solution.
The cells were fixed with 4.0\% paraformaldehyde for 30 min. 
The nuclei of the ND-HeLa cells were then stained with Hoechst 33342 solution for one hour.
After these staining treatments, the cells were washed two times using the PBS solution. Finally, the cells were soaked with PBS solution.
The stained ND-HeLa cells were observed for obtaining the three-dimensional distribution of NDs with a using a super-resolution structured illumination microscope (N-SIM; Nikon). 
The excitation and emission wavelengths were 325/400-450 nm and 550/600-700 nm for nucleus and NDs, respectively.

\subsection{Cell preparation and ND labeling for ODMR measurements}
The HeLa cells were incubated with the NDs (10 $\si{\ug}$/mL) in home-built antenna-integrated glass-based dishes in a transduction medium (E-MEM containing 10\% FBS and 100 U/mL penicillin/streptomycin) at 37 $\si{\degreeCelsius}$. 
After a 24-h incubation, the cells were washed three times using the PBS solution and filled with the transduction medium without phenol red. 
The phenol red was removed to avoid heat generation upon the absorption of green laser light.

\subsection{Quantum thermometry}
\begin{figure*}[th!]
\centering
\includegraphics[width=13cm]{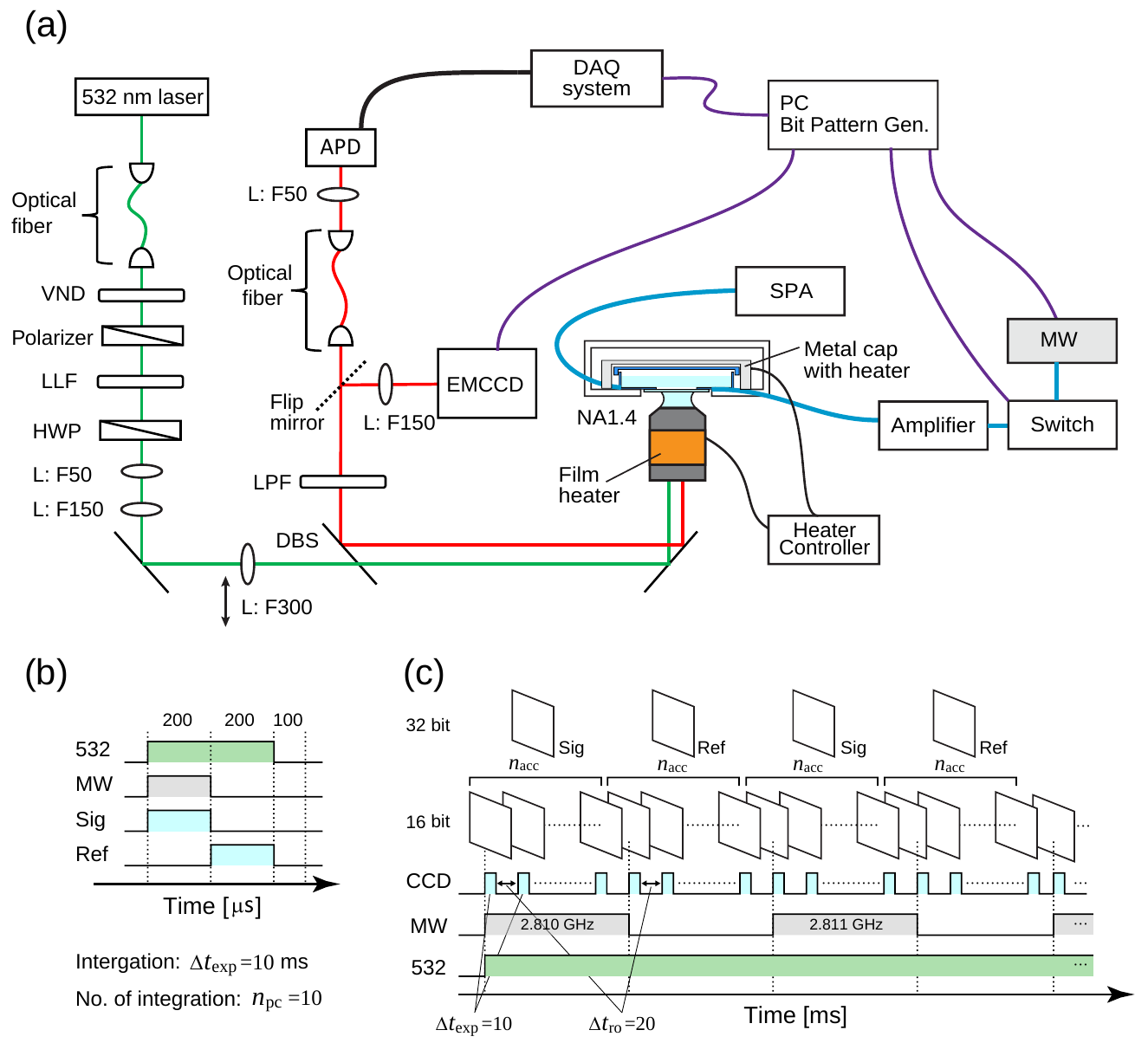}
\caption{(a) Schematic of the ODMR setup including the optical microscope and microwave spin resonance systems. VND: variable neutral density filter. LLF: laser-line filter. HWP: half-wave plate. L: lens. DBS: dichroic beam splitter. LPF: long-pass filter. EMCCD: electron multiplying charge-coupled device camera. APD: avalanche photodiode. SPA: spectrum analyzer. MW: microwave source. DAQ: data-acquisition board. 
Pulse control sequences for the (b) confocal detection and (c) wide-field ODMR detection. Sig: signal. Ref: reference. $\Delta t_{\rm ro}$: readout time, 20 ms. $n_{\rm acc}$: number of image accumulation, 50 or 100. $\Delta t_{\rm exp}$: exposure time of camera, 10 ms. The frequency sweep in the wide-field ODMR started at 2.810 GHz with a step frequency of 1 MHz.}
\label{fig1}
\end{figure*}

We used a home-built quantum thermometry system that was previously described in ~\cite{yukawa2020quantum}, but the detection system was reconstructed to allow both confocal and wide-field ODMR detection (Fig.~\ref{fig1}(a)).
A continuous-wave 532-nm laser was used for optical excitation.
An oil-immersion microscope objective with a numerical aperture of 1.4 was used for the excitation and fluorescence collection. 
The NV fluorescence was filtered with a dichroic beam splitter, and a long-pass filter was used to remove the residual green laser scattering. 
For confocal detection, the fluorescence was coupled to an optical fiber that acted as a pinhole (1550HP, Thorlabs). 
The fiber-coupled fluorescence was detected using an avalanche photodiode (APD) (SPCM AQRH-14, Excelitas). 
For wide-field ODMR detection, the excitation laser was focused on the back focal plane of the objective to illuminate the entire area of the field of view, and the fluorescence was imaged with an electron multiplying charge coupled device (EMCCD) camera (Evolve Delta, Photometrics) via a $f$150-mm achromatic lens. 
In both of the detection methods, the excitation intensity was adjusted to $\sim 10 \ \rm{W} / {\rm cm}^{2}$. 
The sample was mounted on a $xyz$-piezo stage (Piezosystemjena, Tritor-100SG) to enable fine translation and positional scanning.
Note that an external magnetic field was not applied in this study. 

The microwaves were generated from a microwave source (SMB100A, Rohde \& Schwarz) and amplified by a maximum factor of 45 dB (ZHL-16W-43+, Mini-circuit). The microwaves were fed to the linear antenna (25-$\si{\um}$ copper wire) of the home-made cell culture dishes. 
The typical microwave excitation power was estimated to be 10--50 mW (10--17 dBm) at the linear antenna by considering the source output, amplifier gain, and the experimentally determined $S_{21}$ of the antenna system, which provides microwave magnetic field of more than 2--5 gauss in 20 $\si{\um}$ from the antenna.
APD detection was gated for ON and OFF states of microwave irradiation using a radiofrequency switch (ZYSWA-2-50DRS, Mini-circuit) and bit pattern generator (PBESR-PRO-300, Spincore). 
The gate width for the confocal continuous (CW)-ODMR detection was 200 $\si{\us}$, common to both gates, followed by an initialization time of 100 $\si{\us}$, which gave $I_{\rm PL}^{\rm ON}$ and $I_{\rm PL}^{\rm OFF}$ with a frequency of 2 kHz~\cite{yukawa2020quantum} (Fig.~\ref{fig1}(b)). 
At each microwave frequency during the frequency sweep, the ND fluorescence was photon-counted for an integration time of $\Delta t_{\rm pc} = 100$ ms for a single frequency.
The frequency sweep was repeated $n_{\rm pc}$ times to obtain a high signal-to-noise ratio (SNR) in the CW-ODMR spectra.
For wide-field ODMR observation, we fed the camera with trigger pulses from the bit pattern generator to slave its operation (Fig.~\ref{fig1}(c)). 
The camera acquired a 16-bit image with an exposure time of $\Delta t_{\rm exp} = 10$ ms, followed by a readout time of $\Delta t_{\rm exp} = 20$ ms. This was repeated $n_{\rm acc}$ times for accumulation. 
The obtained 16-bit $n_{\rm acc}$ images were summed up into a single 32-bit image to obtain an average (signal image).
The microwave was then turned off to acquire the reference image, which was followed by a change in microwave frequency.
The microwave frequency was scanned with a frequency step of $\Delta f_{\rm step} =$ 1 MHz in the range of $f_{\rm start}$ to $f_{\rm end}$. 
Therefore, we finally obtained a large data set of 32-bit images that contained $2N = 2 (\Delta f_{\rm step})^{-1} |f_{\rm start} - f_{\rm end}|$.

\subsection{Image analysis}
The obtained $2N = 2 (\Delta f_{\rm step})^{-1} |f_{\rm start} - f_{\rm end}|$ image sets were first sorted according to the respective microwave frequencies, and further according to the two types of $N$ image sets for the microwave ON and OFF states. 
To obtain the ODMR contrast, pixel values of the microwave ON images (signal images) were divided by those of the microwave OFF images (reference images). 
This image data processing was performed using the Fiji/ImageJ software~\cite{schindelin2012fiji}.
The region of interest (ROI) determined to extract the fluorescence intensity is described in Sec.~\ref{sec3}.

\subsection{Temperature control of the stage-top incubator}
In the home-built stage-top incubator, we heated the antenna-integrated dishes in two directions, i.e., from the direction of the oil-immersion objective and metal cap (Fig.~\ref{fig1}(a)). 
The temperature of the dish ($T_{\rm d}$) was varied by controlling the temperature of these heat sources that was set to the same temperature ($T_{\rm i}$) using the PID-feedback controller of the foil heater, which wrapped the objective (Thorlabs, HT10K \& TC200, temperature precision: $\pm$ 0.1 $\si{\degreeCelsius}$).
The immersion oil used was Olympus Type-F.
$T_{\rm d}$ was calibrated by inserting a tiny flat Pt100 resistance temperature probe (Netsushin, NFR-CF2-0505-30-100S-1-2000PFA-A-4, $5 \times 5 \times 0.2 \ {\rm mm}^{3}$) in the water media in the dishes, and varying $T_{\rm i}$ while monitoring $T_{\rm d}$.
The temperature probe was read using a high-precision handheld thermometer (WIKA, CTH7000, temperature precision: $\pm$ 0.02 K). 
Consequently, we obtained the following relation, $T_{\rm d} = 5.51 + 0.814 T_{\rm i}$ ($\si{\degreeCelsius}$) according to Fig.~\ref{figS-tcalib}.
Note that $T_{\rm i}$ was monitored directly on top of the foil heater wrapped around the objective and capping metal cover.

\section{Results and Discussion}
\label{sec3}
\subsection{Comparison of wide-field ODMR detection with confocal detection}
\label{sec3.1}
The SNR of the ODMR spectra is dependent on the total photon count measured by the detectors, i.e., the EMCCD camera for the wide-field and APD for the confocal detection. Therefore, we determined the measurement parameters of wide-field detection that provide a comparable SNR in the confocal method.
In wide-field detection, the full-well capacity (FWC) and digital resolution of the analog-to-digital converter (ADC) of the camera strongly limit the measurement strategy.
The FWC of our EMCDD is 185,000 e$^-$ for a single pixel, and the ADC resolution is 16-bit expressing the stored electrons with the range of 0--65,535 counts (cts). 
Considering the quantum efficiency in the photoelectron conversion process near unity based on the manufacturer specification sheet (strictly 90\%), the camera gives a pixel value of 1 for $\sim 3$ photons in a single image.
This pixel value needs to be related to the APD photon count.

\begin{figure}[th!]
\centering
\includegraphics[width=9cm]{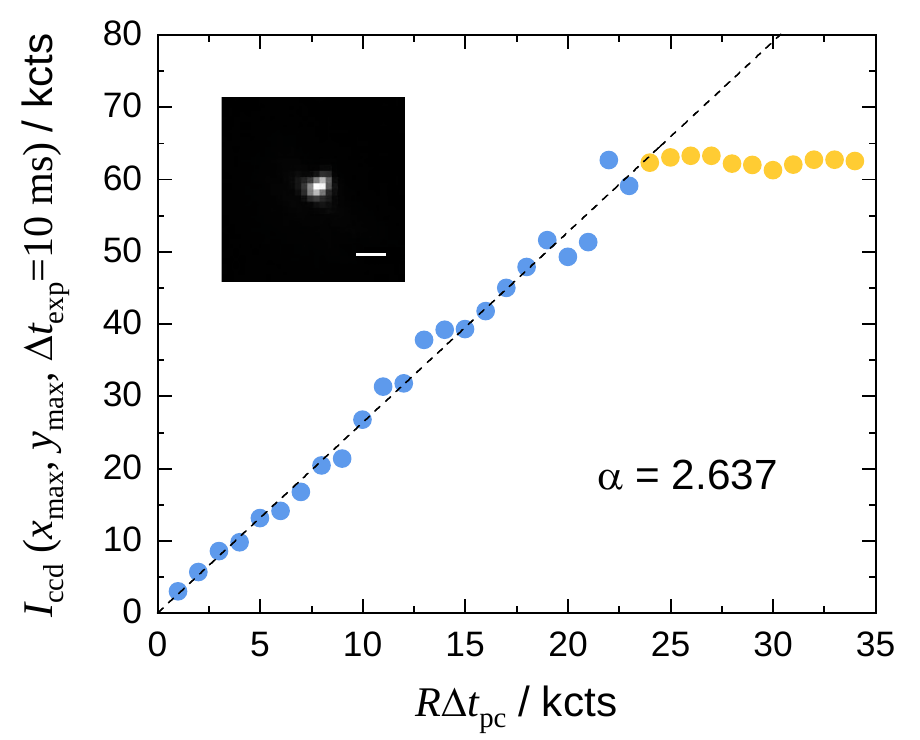}
\caption{Relation between the maximal pixel value of the camera $I_{\rm ccd}(x_{\rm max}, y_{\rm max}, \Delta t_{\rm exp} = 10 \ {\rm ms})$ and the APD photon count during the integration time of $\Delta t_{\rm pc} = 10$ ms. (Inset) wide-field fluorescence image of the target ND. Scale bar: 1 $\si{\um}$.}
\label{fig2-1}
\end{figure}

Figure~\ref{fig2-1}(a) shows such a relation between the pixel value and APD photon count. Specifically, the maximum pixel value of a single ND fluorescence in the ROI for an exposure time of $\Delta t_{\rm exp} = 10$ ms, $I_{\rm ccd}(x_{\rm max}$, $y_{\rm max}$, and $\Delta t_{\rm exp} = 10 \ {\rm ms})$, as a function of the APD photon count during 10 ms ($R \Delta t_{\rm pc}$) in the semi-confocal detection. 
Here, $R$ is the photon count rate (cps: counts per second), semi-confocal means that the ND is excited via epi-illumination, and the fluorescence is detected through the pinhole.
$I_{\rm ccd}(x_{\rm max}, y_{\rm max}, \Delta t_{\rm exp} = 10 \ {\rm ms}$) linearly increases with a slope of $\alpha = 2.637$, and saturates beyond 65,535 cts at the FWC, where $\alpha$ indicates the number of electrons that show a pixel value of one.
Considering the photon count of $\sim 147,000$ cts (converted from the pixel value of 52,000 cts) and the corresponding APD count ($R \Delta t_{\rm pc} = 20,000$ cts), the optical throughput of the pinhole is determined as $\sim$14\%.

With this relation, we can determine the measurement parameters of the wide-field detection based on the confocal ODMR measurements.
In confocal ODMR detection, single measurements ($R \Delta t_{\rm pc}$) for individual frequencies are accumulated $n_{\rm pcc}$ times, which gives the total APD photon count ($I_{\rm APD}$) as follows, 
\begin{equation}
    I_{\rm APD} = n_{\rm pc} R \Delta t_{\rm pc}. 
\end{equation}
Assuming the parameters for the confocal ODMR measurements used in the following Sec.~\label{sec3-1} as $R = 2$ Mcps, $\Delta t_{\rm pc} = 100$ ms, and $n_{\rm pc} =10$ ms, we obtain $I_{\rm APD} = 2$ Mcts. 
Because the SNR is given as $(\sqrt{I_{\rm APD}})^{-1}$, integrating the comparable photon count (or fluorescence flux) in the ROI can provide the same SNR in the wide-field ODMR detection. 
The total fluorescence flux in the ROI ($\tilde{I}_{\rm ccd}$) is given by 
\begin{equation} 
    \tilde{I}_{\rm ccd}\sim 3 n_{\rm acc} \alpha \int_{\rm ROI} I_{\rm ccd}(x, y, \Delta t_{\rm exp}) dxdy, 
\end{equation}
Thus, the wide-field ODMR detection should provide similar SNR to the confocal detection with parameters, $\Delta t_{\rm exp} = 10$ ms and $n_{\rm acc} = 100$.
Note that $\Delta t_{\rm exp} = 10$ ms and $n_{\rm pc} = 50$ are used below, in Sec.~\ref{sec3-2}, which are specified in the captions of each figure.

The temperature sensitivity
can be calculated from the ODMR spectral shape and the detected photon counts~\cite{RevModPhys.92.015004,liu2019coherent}, which vary among NDs because of the material inhomogeneity and microwave-field spatial variation. 
It has a range of 1--2 K$\cdot \sqrt{\rm Hz}$ by considering the linewidth,
ODMR contrast and photon counts to be 10--20 MHz, 0.05--0.1
and 2 Mcps, respectively, which is comparable to previous reports~\cite{yukawa2020quantum,simpson2017non}.

\begin{figure*}[th!]
\centering
\includegraphics[width=12cm]{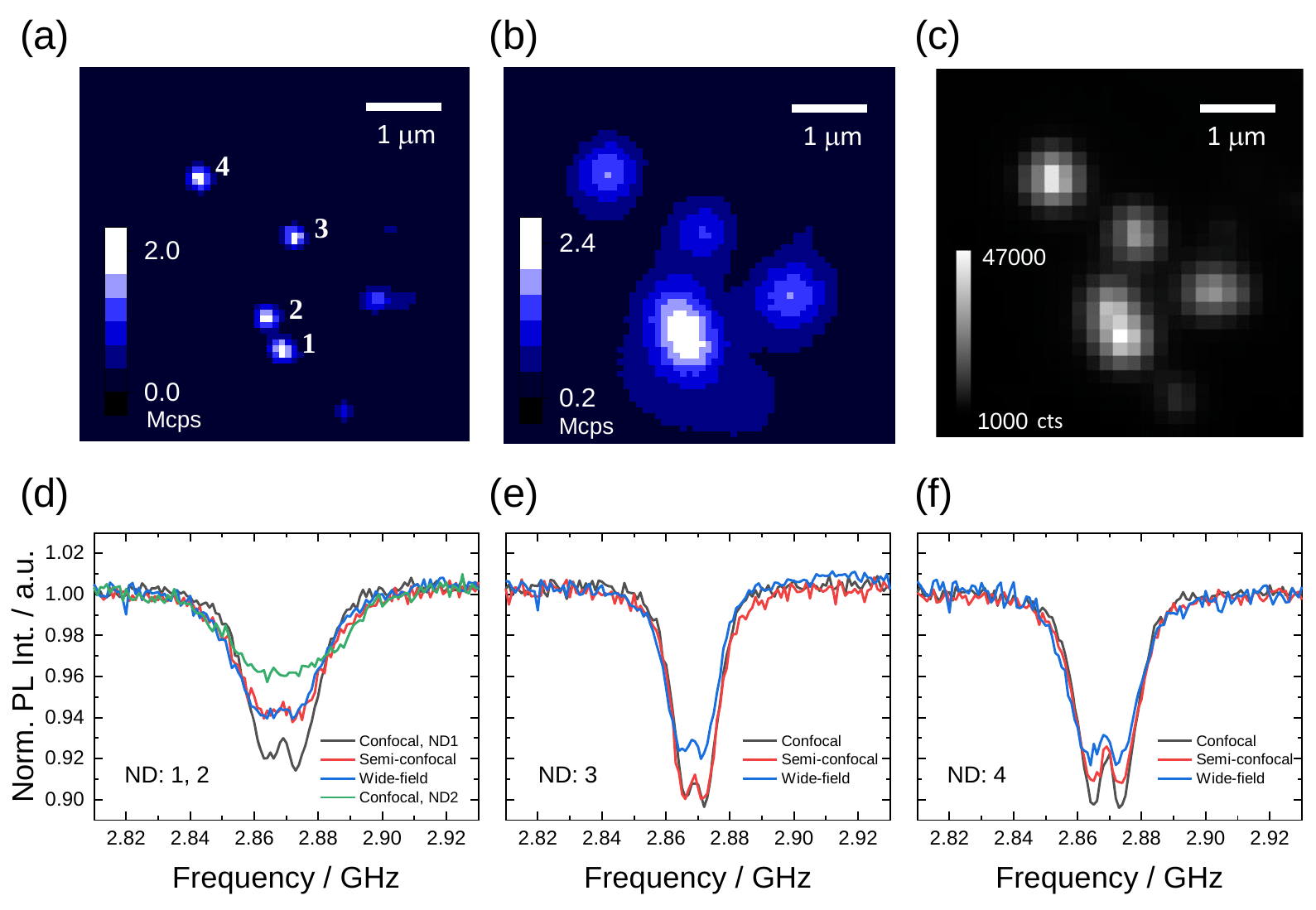}
\caption{Fluorescence images of the NDs measured using (a) confocal scanning, (b) semi-confocal scanning, and (c) wide-field methods. (d) ODMR spectra measured using the three methods for the NDs designated, in Fig.~\ref{fig2-2}(a), as (d) \textbf{1}-\textbf{2}, (e) \textbf{3}, and (f) \textbf{4}. Note that NDs \textbf{1} and \textbf{2} cannot be spatially resolved in the semi-confocal and wide-field detection methods. For the measurements, $n_{\rm acc} = 100$, $\Delta t_{\rm acc} = 10$ ms, $n_{\rm pc} = 10$, and $\Delta t_{\rm pc} = 100$ ms were used.}
\label{fig2-2}
\end{figure*}

Next, we perform ODMR measurements for both the detection methods for the same NDs with comparable photon flux. 
Figures~\ref{fig2-2}(a)--(c) show the fluorescence images of the NDs in confocal scanning (point excitation and pinhole detection), semi-confocal scanning (wide-field excitation and pinhole detection), and wide-field imaging (wide-field excitation and camera detection), respectively.
Confocal scanning provides the best spatial resolution, while semi-confocal scanning and wide-field imaging provide relatively poor resolutions (similar to each other).
Figures~\ref{fig2-2}(d)--(f) show the ODMR spectra measured using the three methods for the NDs designated as \textbf{1}--\textbf{4}, (\textbf{1} and \textbf{2} are not resolvable in Fig.~\ref{fig2-2}(e), (f)), respectively.
In all of the cases, confocal detection provided relatively better ODMR depth by a factor of $\sim 1.3$, compared with those measured in the wide-field detection. 
The difference of the ODMR depth between the detection methods arises from the background contribution of the fluorescence to the ODMR detection.
The background fluorescence is not microwave active, and only acts as the offset of the ODMR spectrum as discussed previously~\cite{fujiwara2016manipulation}.
The semi-confocal detection provided intermediate results compared to the other two methods; the ODMR depth depends on the object shape and size reflecting how much background fluorescence is included. 
It provides similar depths to the wide-field results for NDs \textbf{1} and \textbf{2} because of the large spot size (Fig.~\ref{fig2-2}(d)), while it improves compared to the wide-field (and reaches to the ODMR depth of the confocal results) for NDs \textbf{3} and \textbf{4} owing to their isolated spots (Figs.~\ref{fig2-2}(e) and (f)).
In contrast to the ODMR depth, the ODMR frequency does not vary between the detection methods. This is an important result for quantum thermometry, because the center frequency measured in the different methods should be consistent.

\subsection{Effect of the positional drift of the NDs during wide-field ODMR detection}
\label{sec3-2}
Having understood wide-field ODMR in connection with the confocal method, we subsequently investigate the effects of the positional drift and fluorescence saturation during the measurements in wide-field detection. 
In many biological experiments, positional drift of the NDs frequently occur owing to the dynamic change in structures and locomotion~\cite{fujiwara2019realtime,haziza2017fluorescent}. Such structural changes cause significant variation of optical transmission, thereby increasing or decreasing the fluorescence intensity and sometimes causing pixel saturation.
The saturation of the pixel intensity in the camera affects only the depth and width of the ODMR spectrum, and in principle, does not cause frequency shift.
However, it can cause frequency shift when the drift of the NDs are not properly managed, as described below.

\begin{figure*}[th!]
\centering
\includegraphics[width=12cm]{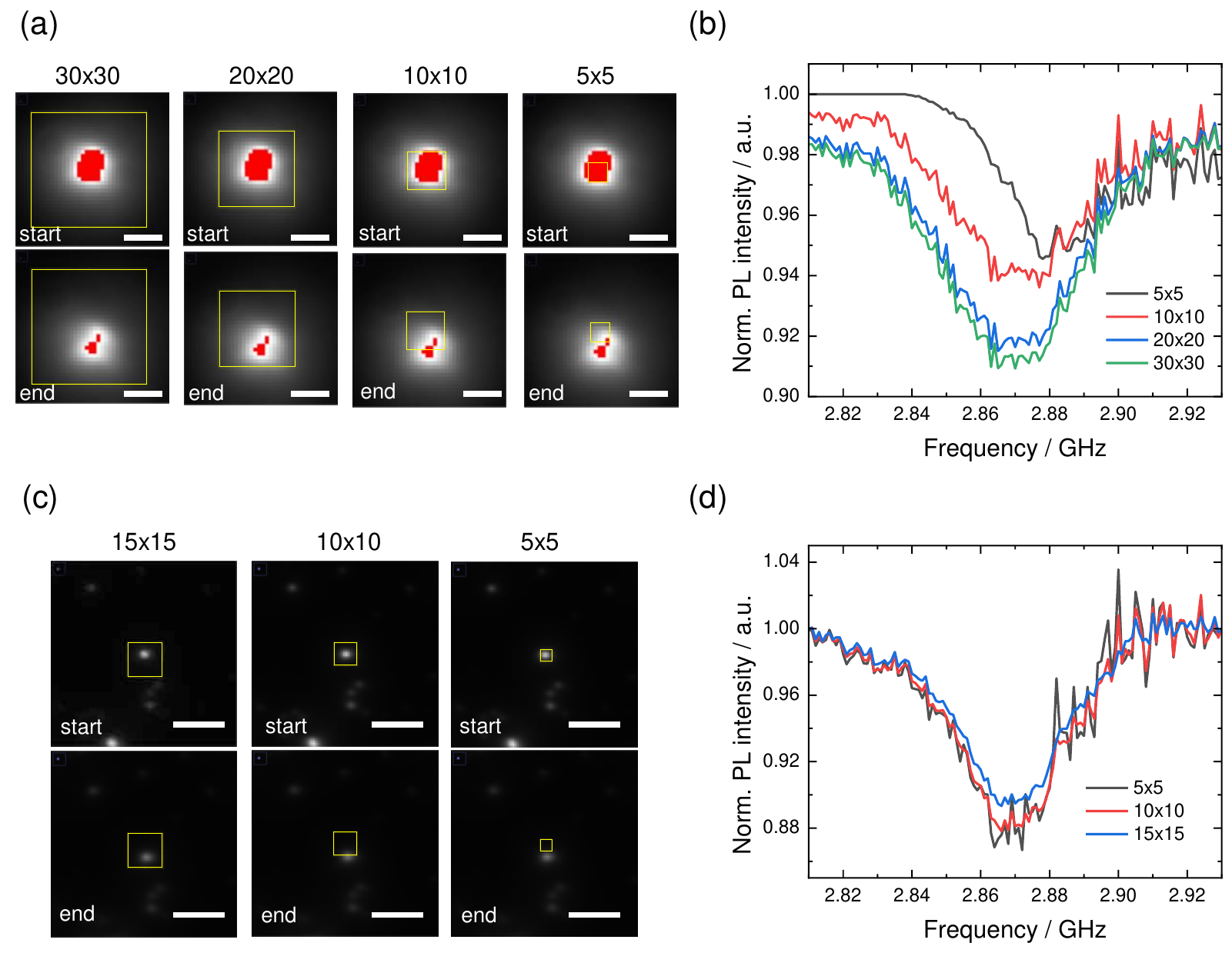}
\caption{(a) Fluorescence images of a single ND with fluorescence saturation at $f_{\rm start} = 2.81$ GHz and $f_{\rm start} = 2.93$ GHz in the frequency sweep with different ROIs ($30 \times 30$, $20 \times 20$, $10 \times 10$, and $5 \times 5$ pixels). Scale bar: 2 $\si{\um}$. (b) The corresponding ODMR spectra of the four types of binning regions. (c) Fluorescence images of a single ND without fluorescence saturation at the start and end of the frequency sweep with different ROIs ($15 \times 15$, $10 \times 10$, and $5 \times 5$ pixels). Scale bar: 5 $\si{\um}$. (d) The corresponding ODMR spectra of the three types of ROIs. For all the experiments, $n_{\rm acc} = 50$ and $\Delta t_{\rm exp} = 10$ ms were used.}
\label{fig3-1}
\end{figure*}

Figures~\ref{fig3-1}(a) and (b) show images of a single ND at $f_{\rm start} = 2.81$ GHz and $f_{\rm end} = 2.93$ GHz of the frequency sweep in the ODMR measurement for different sizes of ROIs and the corresponding ODMR spectra, respectively. 
The ND was drifted in $xyz$ directions during the frequency sweep of approximately 6 min. 
We set four kinds of gradually decreasing ROIs, as $30 \times 30$, $20 \times 20$, $10 \times 10$, and $5 \times 5$ pixels.
As the ROIs become smaller, the drifted ND moves out. 
Accordingly, the ODMR spectra exhibit a decrease in the contrast and shift of the center frequency, which are particularly prominent below $10 \times 10$ pixels. 
Conversely, the spectral shape of the ODMR is not affected by the positional drift when there is no saturation in the ROIs.
Figures~\ref{fig3-1}(c) and (d) show the fluorescence images of a single ND without saturation and the corresponding ODMR spectra, respectively. 
As the ROI decreases, the ODMR spectrum associates significant noise particularly in the higher frequency side as the fluorescence spot cannot stay inside the ROIs because of the positional drift. 

\begin{figure}[th!]
\centering
\includegraphics[width=12cm]{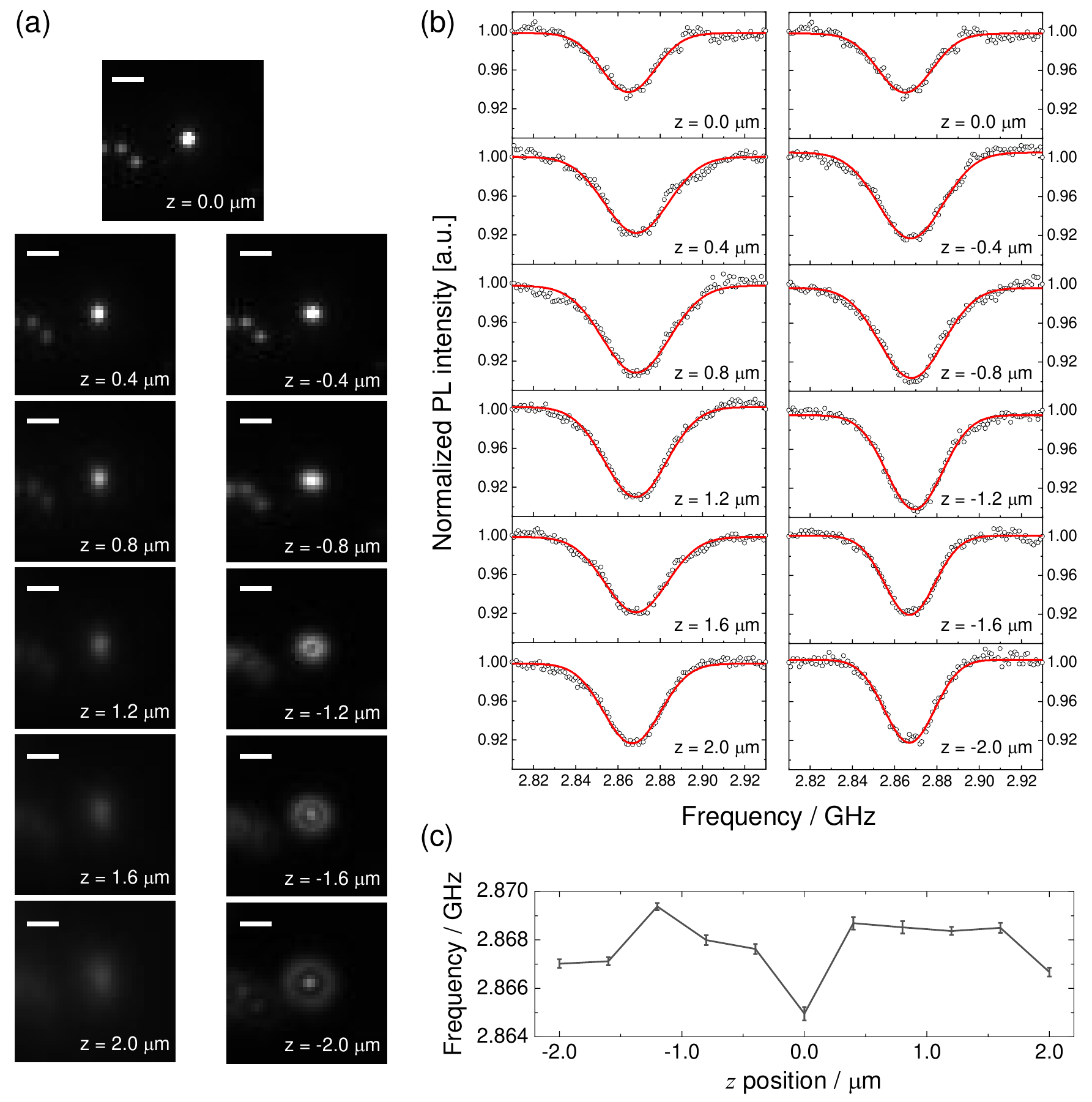}
\caption{(a) Fluorescence images of a single ND with $z$-positional shift of $\pm \ 2 \ \si{\um}$ with respect to the exact focus ($z = 0 \ \si{\um}$) and (b) the corresponding ODMR spectra. Scale bar: 2 $\si{\um}$. (c) The variation of the central frequency of the ODMR spectra. $n_{\rm acc} = 50$ and $\Delta t_{\rm exp} = 10$ ms were used for all the measurements.}
\label{fig3-2}
\end{figure}

Next, the $z$-positional variation of the NDs affecting the wide-field ODMR detection is characterized. We consider this factor because (1) the mechanical distortion of the microscope system and intra-cellular transportation move the NDs in the $z$ axis, and (2) there are number of blurred ND spots (positional variation in the $z$ axis) in a single focal plane owing to the cellular height. 
Figures~\ref{fig3-2} (a) and (b) show a set of fluorescence images of a single ND for the $z$-positional variation of $\pm \ 2 \ \si{\um}$ and the corresponding ODMR spectra, respectively.
As the $z$-position is shifted, the ND is defocused, exhibiting the 3D shape of the point spread function.
Nevertheless, the ODMR spectrum does not vary significantly as long as the ROI is sufficiently large. 
Figure~\ref{fig3-2} (c) shows the center frequency of the ODMR spectrum at each $z$-position determined by Gaussian fitting.
The center frequency exhibits a fluctuation of 370 kHz (s.e.), which is in good agreement with the fluctuation obtained previously for the confocal detection~\cite{fujiwara2019monitoring}. 
Note that the Gaussian that was selected as the fitting function is a practical approach. 
The spectral shape of the NDs have a large variation including Gaussian-like and Lorentzian-like variations~\cite{fujiwara2020arxiv}. 
However, none of these analytic expressions are theoretically correct, because the real formulation needs to consider the interference between the spin states under zero magnetic field, which does not provide simple analytic expressions~\cite{matsuzaki2016optically}. 
Therefore, we use Gaussian in this study to extract the ODMR center frequency. 

The present results regarding the dependence of the wide-field ODMR on the positional variation of the NDs can be summarized as:
(1) The measurement noise (or temperature precision) is dependent on the ND fluorescence flux summed up in the ROI.
(2) The blurred fluorescence spots of individual NDs at a single focal plane does not cause ODMR shift as long as the ROI is properly set.
(3) The pixel saturation of the camera requires particularly careful treatment as it may cause measurement artifacts. 
Increasing the fluorescence intensity and number of measurable NDs in a single image is an important factor in achieving high SNR, but it inevitably saturates the fluorescence of brighter NDs. This is because usually the excitation power is adjusted such that the majority of the NDs gain 40,000--50,000 pixel values. 
Moreover, quantifying this tolerance of the pixel saturation in relation to the ODMR spectra is important in real experiments on biological samples, because it is quite hard to adjust the fluorescence intensity such that it does not to exceed the saturation limit for every single image during thermometric measurements.
It is important to pre-characterize the amount of pixel saturation that is practically acceptable for the temperature measurement.

\subsection{Three dimensional distribution of NDs in HeLa cells}
\begin{figure}[th!]
\centering
\includegraphics[width=10cm]{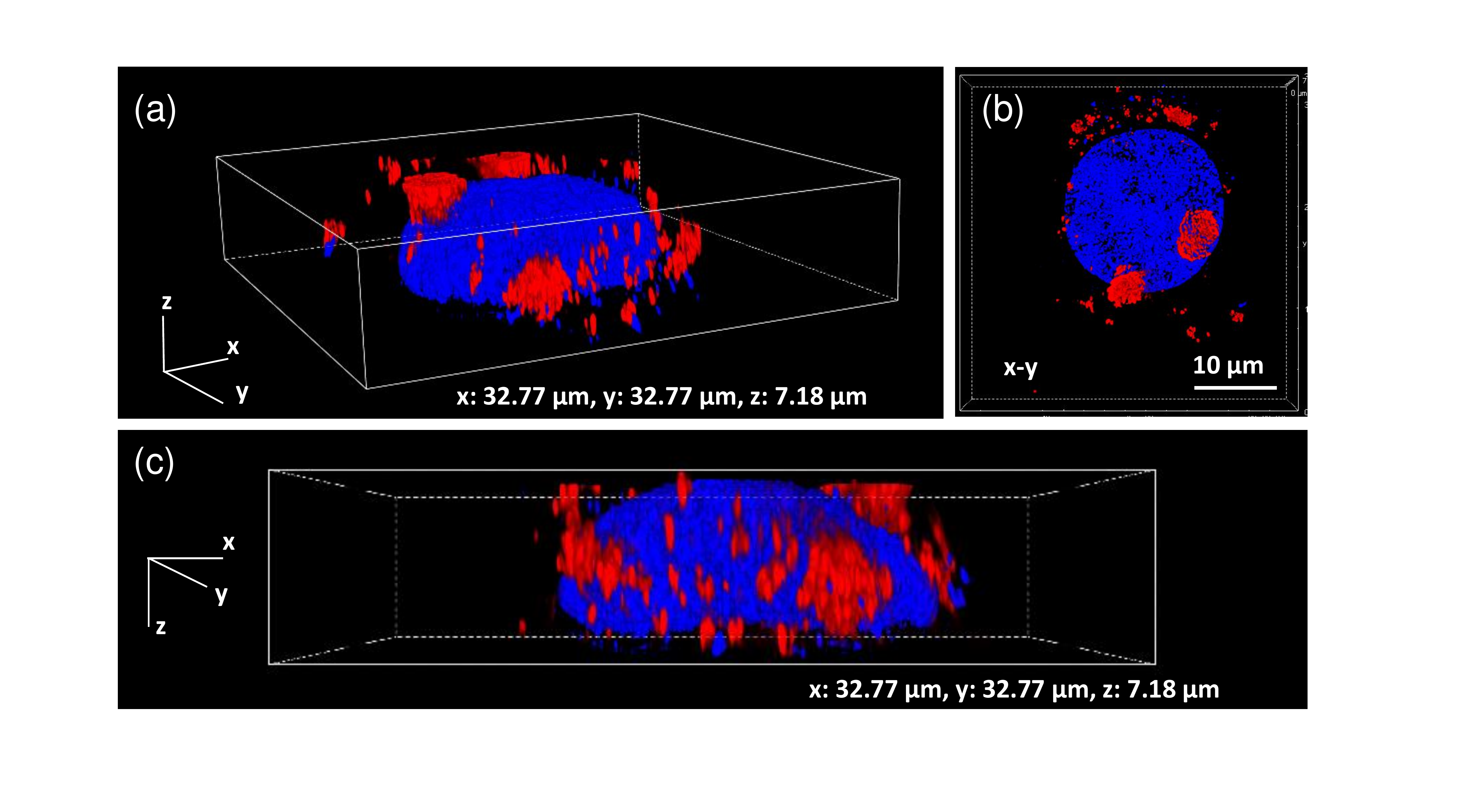}
\caption{Super-resolution fluorescence image of ND-labeled HeLa cells fixed on coverslip.
(a) Three-dimensional fluorescence images of nucleus (blue) and NDs (red). 
(b, c) The corresponding two-dimensional images of $xy$ and $xz$.}
\label{fig4}
\end{figure}

To quantify the $z$-positional distribution of NDs in HeLa cells (and their spatial distribution in the $xy$ plane ), we perform super-resolution imaging of ND-labeled HeLa cells that are fixed on a coverslip, as shown in Fig.~\ref{fig4}. 
Typically, HeLa cells have a two dimensional size of $40 \times 40 \  \si{\um}^2$; furthermore, they have a height of $7 \  \si{\um}$ near the nucleus.  
The NDs around the nucleus were uniformly distributed. 
It is important to note that the distribution of NDs in cells significantly depends on cell types. 
For example, adipose-tissue-derived stem cells (ASCs) have significantly flat structures ($30 \times 30 \ \si{\um}^2$ in the $xy$ plane and $2 \  \si{\um}$ in $z$)~\cite{yukawa2020quantum}.
In ASCs, more NDs can be focused in the wide-field fluorescence image in contrast to HeLa cells.

Although there are approximately $200$ NDs in the cells, not all NDs can be used for the quantum thermometry because (1) some NDs (approximately $20$--$30$\%) show significantly broad spectral lines that cannot be used for quantum thermometry (particularly, peak-shift detection is difficult) and (2) practically, only $10$--$20$\% of the NDs can be focused in the tolerable focusing range in the z-direction because of the substantial height of cells  in the $z$ axis compared to the focal depth.
In the subsequent experiments involving wide-field ODMR measurements in living ND-labeled HeLa cells, $20$--$30$ NDs were predominantly available for measurements.
Note that the number of NDs inside cells can be increased by employing a higher ND concentration.
The selection of ND concentration is dependent on the purpose.
In our experiment, we used a relatively low concentration to focus on individual NDs so that the principles of the system, with regard to wide-field ODMR detection relative to quantum thermometry, could be investigated. 
Note also that adding a $z$-scanning function, namely a $3D$-volumetric wide-field ODMR measurement, can be a candidate for increasing the number of NDs available in future experiments.

\subsection{Wide-field ODMR in HeLa cells}
\begin{figure}[th!]
\centering
\includegraphics[width=13cm]{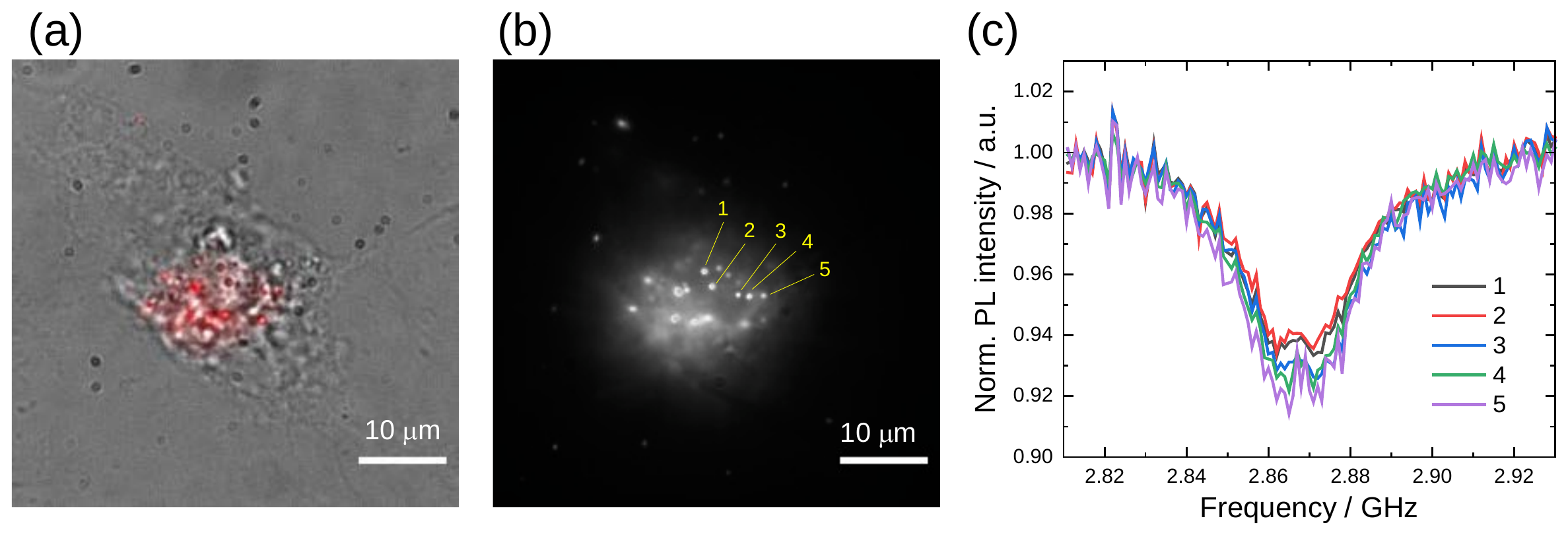}
\caption{Microscope images of living ND-labeled HeLa cells; (a) merged (bright field and red fluorescence) and (b) red fluorescence only. (c) ODMR spectra of NDs \textbf{1}--\textbf{5}. $n_{\rm acc} = 50$ and $\Delta t_{\rm exp} = 10$ ms were used for all the measurements.}
\label{fig5}
\end{figure}

Having understood the system principles of wide-field detection in detail and its connection to confocal detection, we apply it to multiple NDs in living HeLa cells. 
Figures~\ref{fig5}(a) and (b) show merged (bright-field and red) and red fluorescence images of the ND-labeled HeLa cells, respectively.
In these images, we observed $\sim 20$ clear spots of NDs, and a total of $\sim 50$ NDs were discernible at a single focal plane.
We then performed wide-field ODMR measurement and obtained the ODMR spectrum of multiple NDs by properly setting the ROIs.
Figure~\ref{fig5} (c) shows the representative ODMR spectra of the NDs designated as \textbf{1}--\textbf{5} in Fig.~\ref{fig5}(b). 
Notably, we have obtained similar ODMR spectra for multiple NDs in the wide-field detection.

It should be noted that NDs \textbf{1}, \textbf{2}, and \textbf{3}, are somewhat fluorescence saturated and slightly doughnut shaped due to the electron overflow in the pixel arrays.
However, we can obtain ODMR spectra similar to other NDs. 
Moreover, the fluorescence spots of the NDs are sometimes merged or overlapped during the measurements, because of the intra-cellular transportation and merger of endosomes that encapsulate the NDs. However, these events did not occur in the experiment shown in Fig.~\ref{fig4}. 
The influence of such dynamic mergers of the ND fluorescent spots on the ODMR spectra require further investigation, because merged NDs sometimes exhibit different ODMR spectral shapes.

\subsection{Intracellular temperature measurements in HeLa cells}
\begin{figure}[th!]
\centering
\includegraphics[width=12cm]{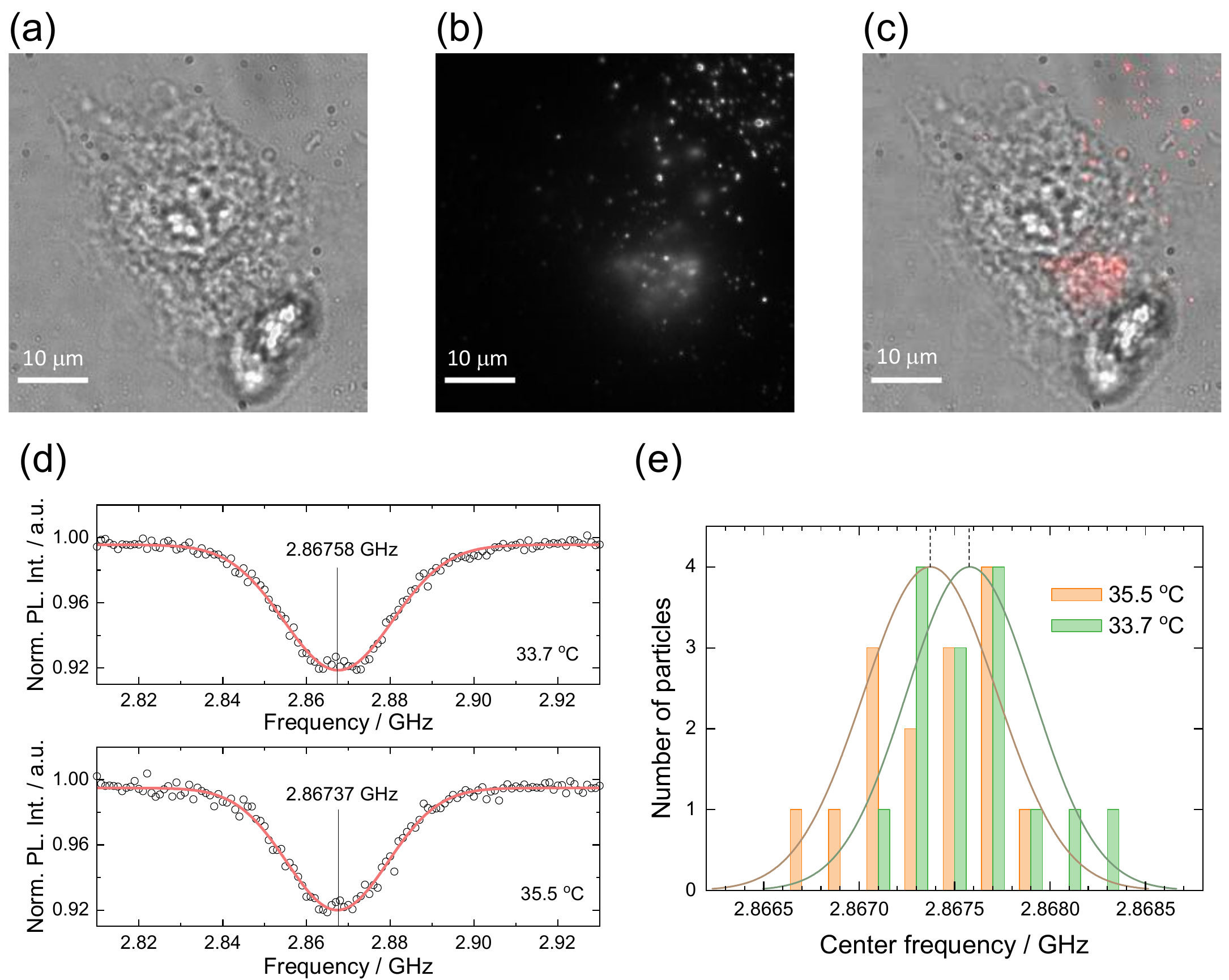}
\caption{Microscope images of the ND-labeled HeLa cells; (a) bright field (b) red fluorescence, and (c) merged. Scale bar: 10 $\si{\um}$. (d) ODMR spectra averaged over 15 NDs in the cell at $T_{\rm d} =$ 35.5 (bottom) and 33.7 $\si{\degreeCelsius}$ (top). (e) Histograms of the center frequency of 15 NDs at the two temperatures with normal distribution fitting. $n_{\rm acc} = 50$ and $\Delta t_{\rm exp} = 10$ ms were used for all the measurements.}
\label{fig6}
\end{figure}

We then performed intra-cellular temperature sensing of living HeLa cells in the present wide-field detection. 
Figures~\ref{fig6}(a)--(c) show the bright-field, red-fluorescence, and merged images of the ND-labelled HeLa cells at 35.5 $\si{\degreeCelsius}$, respectively.
The dish temperature, $T_{\rm d}$, is then varied from 35.5 to 33.7 $\si{\degreeCelsius}$, and the ODMR spectra of the NDs are measured using the wide-field detection at each temperature. %
Figure~\ref{fig6}(d) shows the ODMR spectra averaged over 15 ND fluorescence spots inside the cell at these temperatures.
The center frequency of the ODMR spectrum is shifted by 210 kHz when the dish temperature is varied by -1.8 $\si{\degreeCelsius}$.
The fitting errors of the Gaussian function to the ODMR spectra at $T_{\rm d} =$ 35.5 and 33.7 $\si{\degreeCelsius}$ are 135 and 165 kHz, respectively.

To further confirm the ODMR shift due to the temperature change, we statistically analyzed the temperatures of individual NDs.
Figure~\ref{fig6}(e) shows the histograms of the center frequency of 15 NDs at the two temperatures, where the center frequency is determined by Gaussian fitting.
The ND temperature indication shows a normal distribution, and provides center frequencies of 2.86737 and 2.86758 GHz at $T_{\rm d} =$ 35.5 and 33.7 $\si{\degreeCelsius}$, respectively.
The standard errors of the center frequencies are 86 and 91 kHz, respectively. 
Considering the reported temperature dependence of the zero-field splitting as -74 kHz/K~\cite{PhysRevLett.104.070801,kucsko2013nanometre}, the temperature change of the entire cell sensed by multiple NDs can be calculated as - 2.8 K with an uncertainty of $\sim 1.2$ K.
Thus, our wide-field ND quantum thermometry has succeeded in measuring the intra-cellular temperature change.

\subsection{Prospective to extend to real-time measurements for thermal live-imaging of cells}
The present study has clarified that wide-field ODMR detection can provide comparable results with confocal detection, as long as the pixel saturation is properly treated. 
The selection of ROIs and the fluorescence brightness of the NDs that lead to pixel saturation are critical points when implementing wide-field ODMR detection in biological thermometry. 
This technical point is camera dependent, where higher bit depths and larger FWCs are required for the camera, as reported for magnetometry applications~\cite{wojciechowski2018contributed}. 
For example, single photon counting cameras~\cite{Poland:15} or neural imaging cameras~\cite{doi:10.1113/jphysiol.2011.219014} may be useful for such applications. 
While wide-field detection is robust against the $z$-positional variation of the NDs, a coarse positional tracking is necessary to obtain full-3D volumetric temperature measurements of entire cells. 
Combining such $z$-positional information with phase imaging techniques~\cite{doi:10.1098/rsos.191921} can provide more precise information of cellular structures, which could aid sub-cellular temperature mapping~\cite{okabe2012intracellular,kiyonaka2013genetically}.
Additionally, the reduction of the measurement times of the ODMR spectra is important. 
For example, the fast-timing image detection mode of the camera may speed up this process. 
Integrating wide-field detection with multi-point ODMR methods seems necessary to improve the temperature precision, as confocal detection has recently demonstrated biological thermometry in living nematode worms upon integration with multi-point ODMR~\cite{fujiwara2019realtime,choi2019}. 
For such purposes, the quantitative relation between confocal and wide-field ODMR detection determined in this study is applicable.

In addition to the development of these measurement techniques, the spatial distribution of NDs strongly affects the measurement strategy.
In the present study, the NDs were freely distributed in the cells. 
In such cases, one can obtain the temperature information of whole cells or certain regions of cells as reported previously~\cite{choi2019,doi:10.1021/jp066387v,yukawa2020quantum,fujiwara2019realtime}.
One such example is the recent demonstration of quantum thermometry of sub-cellular temperature gradient in embryos~\cite{choi2019}, where the spatial distribution of temperature inside embryos is measured using freely distributed NDs while controlling the local temperature. 
With site-specific ND labeling techniques, one may control this ND spatial distribution more effectively.
There are various types of biomolecular conjugation of NDs that enable site-specific temperature measurements of various organelles, such as mitochondria~\cite{doi:10.1021/acsami.6b15954}, cellular membrane~\cite{doi:10.1021/jacs.0c01191}, and lysosomes~\cite{su2017fluorescent}. 
Intra-cellular quantum thermometry thus requires development of both measurement techniques and ND labeling techniques for target biological phenomena.

\section{Conclusion}
In this study, we have analyzed wide-field ND quantum thermometry in detail for application in intra-cellular temperature measurements.
The ODMR spectra obtained using confocal and wide-field detection were compared for the same NDs, and we found that the ODMR was deeper in the confocal detection than in the wide-field detection. This difference of the ODMR depth was because of the different amounts of background fluorescence, but the ODMR center frequency was not significantly different between the two methods. 
The effect of pixel saturation and ND positional drift on wide-field ODMR detection was studied to ascertain the proper determination of the ROIs.
Using this basic system characterization, we performed simultaneous ODMR measurement of multiple NDs in living HeLa cells under temperature changes, and demonstrated that the temperature precision in the wide-field detection was the same as that in the confocal detection.
Furthermore, the future prospects regarding the development of faster and more precise wide-field ODMR temperature detection were discussed.
Our results are significant for the development real-time large-area biological quantum thermometry.

\section*{Funding}
This work was supported in part by the Osaka City University Strategic Research Grant 2017--2019 (M.F., T.M., H.Y., Y.S.) and JSPS-KAKENHI (M.F., H.Y.: 16K13646. Y.N.: 19K15422. M.F.: 17H02741, 19K21935, 20H00335. H.Y.: 17H02731)
M.F. acknowledges the funding from MEXT-LEADER program, Murata Science Foundation, Sumitomo Research Foundation, and Watanabe Foundation.

\section*{Disclosures}
The authors declare no conflicts of interest.

\section*{Appendix}
\renewcommand{\thefigure}{S\arabic{figure}}
\setcounter{figure}{0}

\begin{figure}[th!]
\centering
\includegraphics[width=9cm]{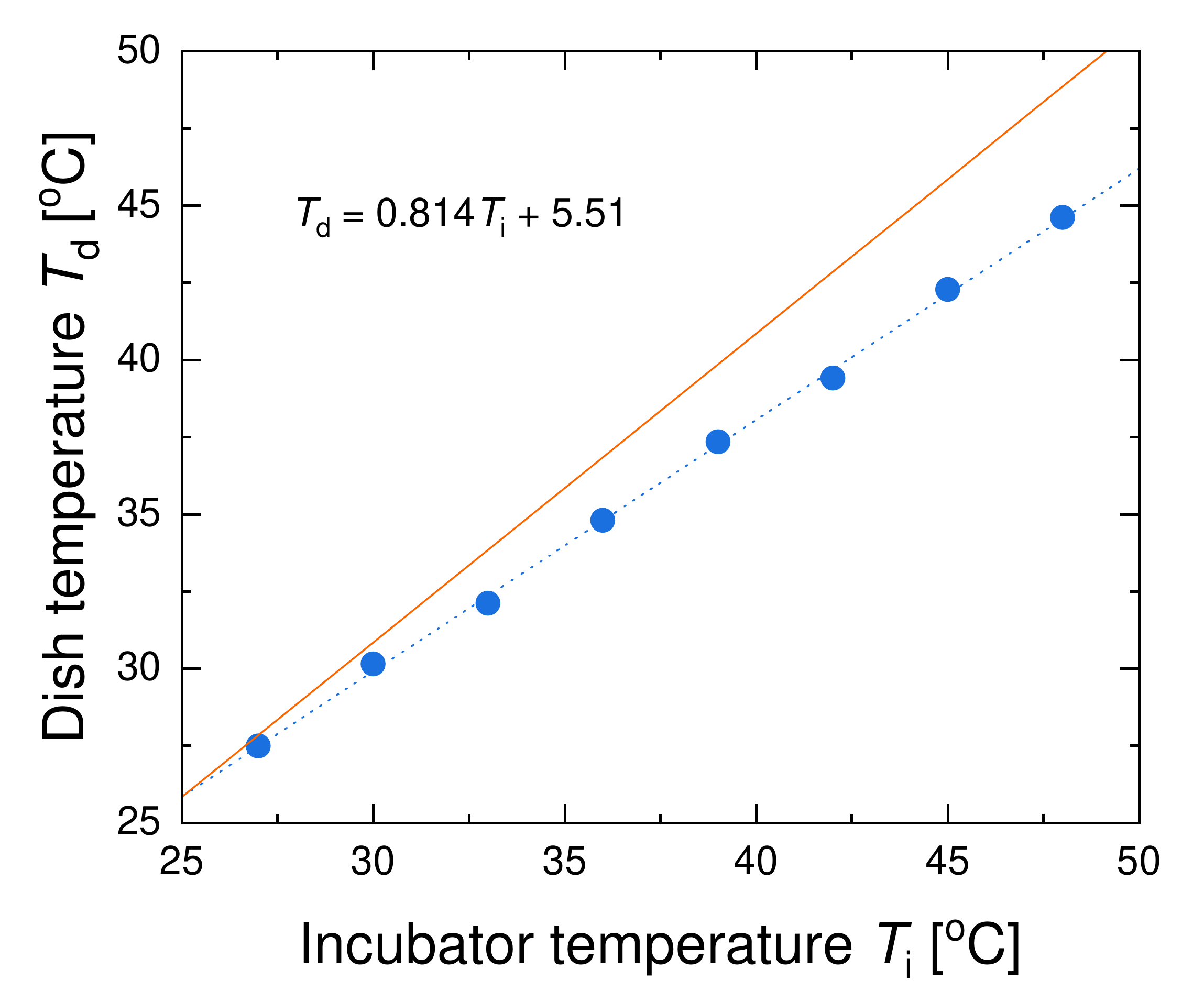}
\caption{Calibration of the dish temperature ($T_{\rm d}$) with the incubator heat source ($T_{\rm i}$). $T_{\rm S} = 5.51 + 0.814 T_{\rm i}$ is obtained. The dotted line is the linear fit. The solid line indicates a slope of one.}
\label{figS-tcalib}
\end{figure}

\begin{figure}[th!]
\centering
\includegraphics[width=12cm]{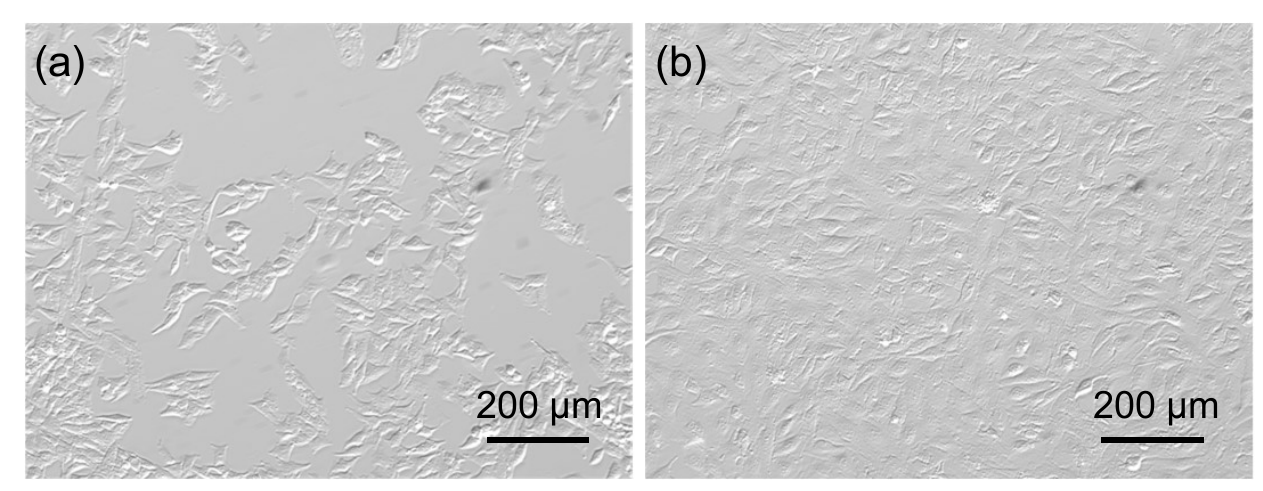}
\caption{Cultured HeLa cells on dishes (a) without and (b) with collagen coating}
\label{figS-collagen}
\end{figure}

%

\end{document}